\documentclass[a4paper]{article}

\usepackage{INTERSPEECH2022}
\usepackage{enumitem}
\usepackage{amsmath}
\usepackage{multirow}

\title{Understanding Audio Features via Trainable Basis Functions}
\name{Kwan Yee Heung$^1$, Kin Wai Cheuk$^{1,2}$, Dorien Herremans$^1$}
\address{
  $^1$Singapore University of Technology and Design\\
  $^2$Agency for Science, Technology and Research}
\email{\{kwanyee\_heung,dorien\_herremans\}@sutd.edu.sg, kinwai\_cheuk@mymail.sutd.edu.sg}

\begin{document}

\maketitle
\begin{abstract}

In this paper we explore the possibility of maximizing the information represented in spectrograms by making the spectrogram basis functions trainable. We experiment with two different tasks, namely keyword spotting (KWS) and automatic speech recognition (ASR). For most neural network models, the architecture and hyperparameters are typically fine-tuned and optimized in experiments. Input features, however, are often treated as fixed. In the case of audio, signals can be mainly expressed in two main ways: raw waveforms (time-domain) or spectrograms (time-frequency-domain). In addition, different spectrogram types are often used and tailored to fit different applications. In our experiments, we allow for this tailoring directly as part of the network. 

Our experimental results show that using trainable basis functions can boost the accuracy of Keyword Spotting (KWS) by 14.2 percentage points, and lower the  Phone Error Rate (PER) by 9.5 percentage points. Although models using trainable basis functions become less effective as the model complexity increases, the trained filter shapes could still provide us with insights on which frequency bins are important for that specific task. From our experiments, we can conclude that trainable basis functions are a useful tool to boost the performance when the model complexity is limited.
\end{abstract}

\section{Introduction}
Feature engineering has been a crucial part of developing a good machine learning model. A range of different audio front-end toolkits have been developed for this purpose in the field of audio~\cite{eyben2010opensmile, eyben2009openear,mcfee2015librosa,bogdanov2013essentia,madmom}. From choosing which features to include~\cite{LI2021328, cheuk2020impact} to allowing the model to learn its own features~\cite{lee2018samplecnn,thickstun2017learning,9427206}, researchers have been trying to extract as much useful information from audio clips as possible. Choosing the most important features from audio signals is more difficult than from images, since audio features can be expressed in either time-domain or time-frequency-domain. Even more so, there exists more than one time-frequency-domain feature representation, for instance, short-time Fourier transform (STFT), constant-Q transform (CQT)~\cite{brown1991calculation}, and Mel-frequency cepstral coefficients (MFCC)~\cite{tiwari2010mfcc}.
Due to the effort required to select important features, some researchers simply use the same spectrogram configuration as in related literature~\cite{berg2021keyword,majumdar2020matchboxnet,10.1007/978-3-030-87802-3_69,2018arXiv180808929C}, hoping that a deep enough model could extract all the information it needs for the task regardless of what audio features are fed to the model. 

nnAudio~\cite{cheuk2020nnaudio} is a trainable front-end tool which allows the optimization of raw waveforms (time-domain) to spectrograms (time-frequency-domain) conversion. This is achieved by implementing the STFT and Mel basis functions as a first layer of the neural network (front-end layer), which allows models to back-propagate the gradient to the front-end. And hence, a custom spectrogram can be `calculated' that suits the task better.

Similar to nnAudio, FastAudio~\cite{fu2021fastaudio}, is also a trainable front-end tool but with shape constraint applied to Mel bases. In their spoofing detection model, the trainable Mel bases results in a higher accuracy than using conventional spectrograms.
Yet, spoofing detection is a binary classification~\cite{balamurali2019toward,wu2017asvspoof,zeinali2019detecting}, and it is still unclear if the same applies to other tasks. In this paper, we extend the idea of trainable basis functions to more tasks such as keyword spotting (KWS) and automatic speech recognition (ASR). More specifically, we want to demonstrate the following two points: 1) Under what circumstances are trainable basis functions useful? 2)
What basis functions are learned after training and why are they better then the original one?

\begin{figure*}[t]
  \centering
  \includegraphics[width=\textwidth]{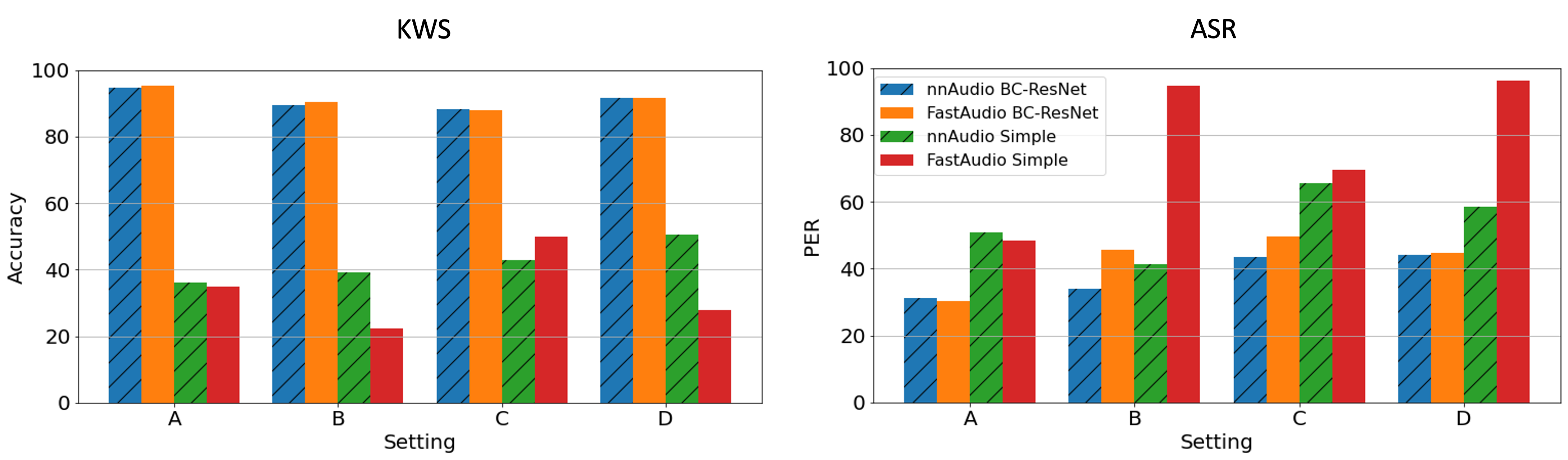}
  \caption{Model performances under four different settings (A-D defined in Section~\ref{sec:exp}) on both KWS and ASR.}
  \label{fig:results}
\end{figure*}

\begin{table*}[]
  \caption{Keyword spotting accuracy when different number of Mel bases are used under four different trainable settings A,B,C,D. (defined in Section~\ref{sec:exp})}
  \centering
  \begin{tabular}{c|ccccc|ccccc}
    \toprule
   {\textbf{\# Mel}} & \multicolumn{5}{c|}{\textbf{nnAudio}}                     & \multicolumn{5}{c}{\textbf{FastAudio}}                   \\
    \midrule
   & A & B & C & D & max. imp.& A & B & C & D & max. imp.\\
    \midrule
10 & 23.9\%   & 37.7\%   & \textbf{39.6\%}    & 34.5\%          &  15.7 ppt.  & 35.0\%   & 8.8\%    & \textbf{36.3\%}             & 17.2\%    & 1.3 ppt.\\
20 & 27.9\%   & 37.4\%   & 32.7\%             & \textbf{38.6\%} &  10.7 ppt.  & 31.8\%   & 19.1\%   & \textbf{46.2\%}    & 24.6\%    & 14.4 ppt.\\
30 & 26.4\%   & 39.7\%   & \textbf{43.6\%}    & 40.4\%          &  14.0 ppt.    & 32.2\%   & 17.7\%   & \textbf{50.0\%}    & 23.7\%    & 17.8 ppt.\\
40 & 36.2\%   & 39.3\%   & 42.8\%             & \textbf{50.4\%} &  14.2 ppt.  & 35.0\%   & 22.3\%   & \textbf{49.8\%}             & 27.8\%    & 14.8 ppt.\\
    \bottomrule
  \end{tabular}
  
\end{table*}

\section{Setup}
\label{sec:exp}
To address the research questions above, a number of experiments are conducted in which we compare the performance of trainable short-time Fourier transform (STFT) $g_\text{STFT}$ and Mel basis functions $g_\text{Mel}$ provided by FastAudio and nnAudio on two tasks: KWS and ASR. We use a broadcasting-residual network (BC-ResNet)~\cite{kim2021broadcasted} as well as a Simple model (constructed with a linear layer) for these two tasks.

nnAudio can produce the same STFT outputs as other major audio processing libraries such as librosa~\cite{mcfee2015librosa} and torchaudio~\cite{yang2021torchaudio}, however, it allows $g_\text{STFT}$ to be trained together with the model, the effect of which we aim to study.

\subsection{Trainable basis functions}
For both nnAudio~\cite{cheuk2020nnaudio} and FastAudio~\cite{fu2021fastaudio}, 40 Mel bases ranging from 0Hz to 8,000Hz are used. We denote the audio front-end as $g$, to which we apply both $g_\text{STFT}$ as well as $g_\text{Mel}$ on the raw waveform $X_t \in [-1,1]^{T_t}$, resulting in a spectrogram $X_f \in [0,+\infty)^{T_f\times 40}$:

\begin{equation}
    \begin{cases}
     g = g_\text{Mel} \circ g_\text{STFT}\\
     X_f = g(X_t)\\
    \end{cases}\,,
\end{equation}
where $T_t$ and $T_f$ are the lengths of the sample in the time- and frequency-domain respectively.

Both nnAudio and FastAudio initialise the $g_\text{Mel}$ as triangular shapes~\cite{Stevens1937ASF, Stevens1940TheRO, fant1949analys, koening1949new}. One major difference between nnAudio and FastAudio is that all Mel bases $g_\text{Mel}$ in FastAudio have the same amplitude, and the bases remain triangular shape in both the initial and training state. In nnAudio on the other hand, the amplitude of the initialised $g_\text{Mel}$ decays as the frequency goes up. During the training state, nnAudio allows $g_\text{Mel}$ to be updated to any shape and with any amplitude ranging from 0 to 1. In our experiments, we explore four different training settings:
\begin{enumerate}[label=\textbf{\Alph*}]
\setlength{\itemsep}{0pt}%
\setlength{\topsep}{0pt} 
\item Both $g_\text{Mel}$ and $g_\text{STFT}$ are non-trainable.
\item $g_\text{Mel}$ is trainable while $g_\text{STFT}$ is fixed.
\item $g_\text{Mel}$ is fixed while $g_\text{STFT}$ is trainable.
\item Both $g_\text{Mel}$ and $g_\text{STFT}$ are trainable.
\end{enumerate}

\subsection{Preprocessing and model architecture}
 The same audio preprocessing pipeline is used throughout our experiments. All of the audio clips are downsampled into 16kHz. All experiments share the same STFT front-end $g_\text{STFT}$  provided by nnAudio with a Hann window size of 480 samples, and a hop size of 160 samples. For $g_\text{Mel}$, 40 Mel bases are used.

For each setting, we consider two model architectures, BC-ResNet~\cite{kim2021broadcasted} and Simple, as the classifier which is denoted by $f$. Hence, the output of the classifier $Y\in [0,1]^C$ can be obtained via:
\begin{equation}
\begin{split}
     Y &= (f \circ g)(X_t)\\
     & = f(X_f)    
\end{split}\,,
\end{equation}
where C is the number of classes depending on the task defined in Section~\ref{tab:KWS} and Section~\ref{tab:ASR}.

\section{Experiments}
\subsection{Keyword Spotting}
\label{tab:KWS}
We use the Google speech commands dataset v2~\cite{warden2018speech} to examine the effects of trainable basis functions. There are total 12 (C=12) different classes in this KWS task, and the dataset has total 35 single wordings. Following existing literature~\cite{ kim2021broadcasted,warden2018speech, rybakov2020streaming}, 10 out of 35 words are chosen ( `down', `go', `left', `no', `off', `on', `right', `stop', `up', `yes'). The remaining 25 words are grouped as class `unknown'. A class `silence' is created from background noise as in existing literature~\cite{ kim2021broadcasted,warden2018speech, rybakov2020streaming}. Due to the class imbalance between the `silence' and `unknown' class, we re-balance the training set by adjusting the sampling weight for each class during training~\cite{ kim2021broadcasted,warden2018speech, rybakov2020streaming}.
Softmax is applied to the model output $Y$ and cross entropy minimized using Adam optimizer with a constant learning rate of $1\times 10^{-3}$.  
All models are trained for 200 epochs with a batch size of 100 as it yields the best accuracy.


Figure~\ref{fig:results} shows the speech command prediction accuracy using different models and settings. When a linear layer (Simple) is used for the prediction (green bars), using a trainable $g_\text{STFT}$ and $g_\text{Mel}$ with nnAudio results in a higher accuracy. When both $g_\text{STFT}$ and $g_\text{Mel}$ are trainable (setting D), the accuracy is 14.2 percentage point higher than with fixed $g_\text{STFT}$ and $g_\text{Mel}$ (setting A). Interestingly, the shape constrains imposed by FastAudio (red bars in Figure~\ref{fig:results}) during training did not help in keyword spotting. On the contrary, shape constrains harmed the model performance. We also tried training the FastAudio front-end with different learning rates (from $1\times 10^{-5}$ to $1\times 10^{-3}$) but none of it yields a result as good as nnAudio's. Therefore, the shape constrains might not work well for all speech-related tasks in general.

When BC-ResNet is used, the keyword spotting accuracy reaches over $90\%$, but the trainable basis functions (settings B-D) become less effective. We believe that when the model is deep enough, it is able to learn how to extract useful information out of the spectrograms, even if they are less tuned to the task. At the same time, a complex model is also very sensitive to the change of the input $X_f$, and therefore trainable kernels together with a complex model might potentially confuse the model instead of helping. When a shallow model is used, it does not have enough power to extract useful information out of $X_f$, hence trainable basis functions produce a better $X_f$ so that a shallow model can achieve a better prediction accuracy. We conducted an ablation study to understand what happens when we reduce the number of Mel bases $g_\text{Mel}$ in the spectrograms (see Table 1). In all cases, trainable basis functions (either setting B, C, or D) yield a better result than non-trainable basis functions (setting A). When nnAudio is used as the trainable $g$, it improves the accuracy by at least 10 percentage points (ppt.). Setting C and D on average yield a better model performance. It shows that trainable STFT kernels are able to extract more information out of the raw waveform $X_t$.

The shape constraint imposed by FastAudio does not improve the accuracy for this task. Whenever the $g_\text{Mel}$ are updated with the shape constraint (setting B and D), the performance becomes worse than when keeping the basis functions non-trainable. Nonetheless, it is interesting to see that with trainable $g_\text{STFT}$ (setting C) the model performance is relatively unchanged when the number of Mel bases in $g_\text{Mel}$ is reduced from 40 to 30.

\subsection{Automatic Speech Recognition}
\label{tab:ASR}

The TIMIT dataset~\cite{garofolo1993timit} is used to study the effect of trainable basis functions on ASR. TIMIT contains 6,300 sentences from 630 speakers.  We mainly focus on phoneme recognition by using the phoneme labels provided in the dataset. There are 61 distinct phoneme labels with 1 separator class, $C=62$ in this task. We measure the model performance using the phone error rate (PER). We use a batch size of 100 to train the model for 400 epochs. The connectionist temporal classification (CTC) loss~\cite{graves2006connectionist} is minimized using an Adam optimizer with a constant learning rate of $1\times 10^{-3}$. As ASR is a much more difficult task than KWS, using a linear layer alone does not perform well. Therefore we use a long short-term memory (LSTM) layer~\cite{hochreiter1997long} together with either a modified BC-ResNet~\cite{kim2021broadcasted} or a linear layer as the classifier. To preserve the time dimension, we modify the BC-ResNet by removing the average pooling along the time dimension such that the output of BC-ResNet has the same number of timesteps $T_f$ as the input spectrogram $X_f$.

Similar to KWS, we only observe improvements when the model is a relatively simple (LSTM + linear layer). When BC-ResNet is used instead of a linear layer, using trainable $g_\text{Mel}$ (setting B) worsens the PER from 31.2 to 33.9. Unlike KWS, trainable $g_\text{Mel}$ (setting B) are more effective than trainable $g_\text{STFT}$ (setting C). When nnAudio Simple is used, the former setting improves the PER from 51.0 to 41.4, and the latter worsen the PER to 65.58. We believe that due to the fact that ASR (one input, many predictions) is more complicated than KSW (one input, one prediction). 

\begin{figure*}[h!]
  \centering
  \includegraphics[width=\textwidth]{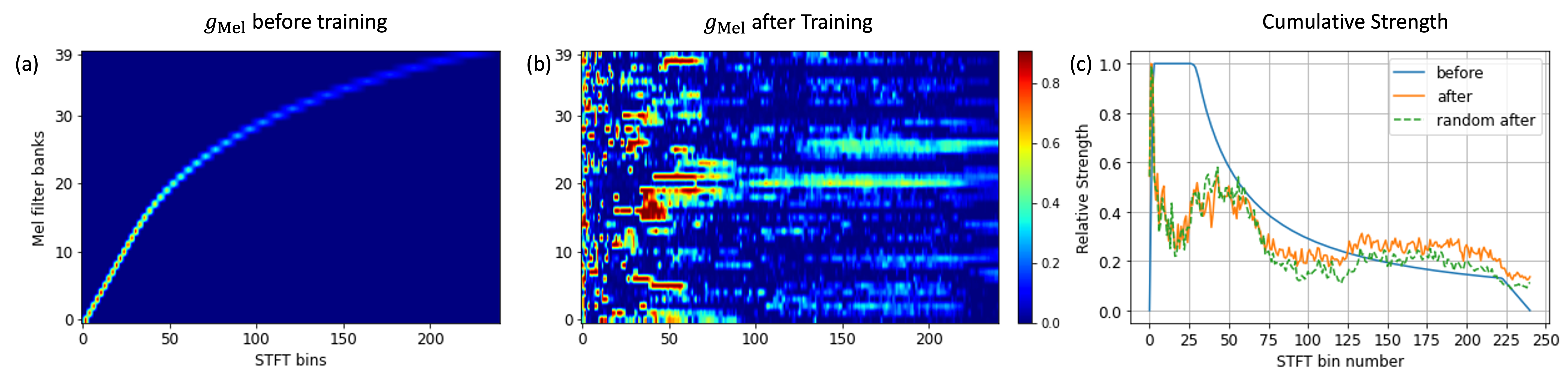}
  \caption{(a)-(b) $g_\text{Mel}$ before and after training on Google speech commands dataset. (c) shows the cumulative sum of different Mel bases in $g_\text{Mel}$.}
  \label{fig:visualization}
\end{figure*}
\begin{figure*}[hbt!]
  \centering
  \includegraphics[width=\textwidth]{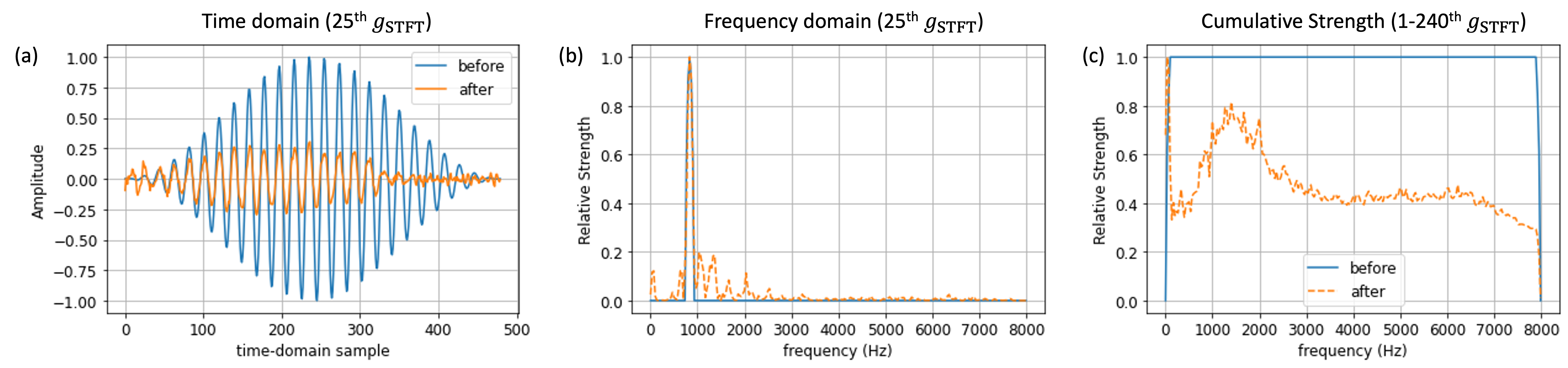}
  \caption{(a)-(b) The 25th STFT basis before and after training (blue and orange lines) on the speech command dataset. (c) Cumulative strengths for 1-240th STFT bases combined.}
  \label{fig:stft_visualization}
\end{figure*}

\begin{figure}[t]
  \centering
  \includegraphics[width=\linewidth]{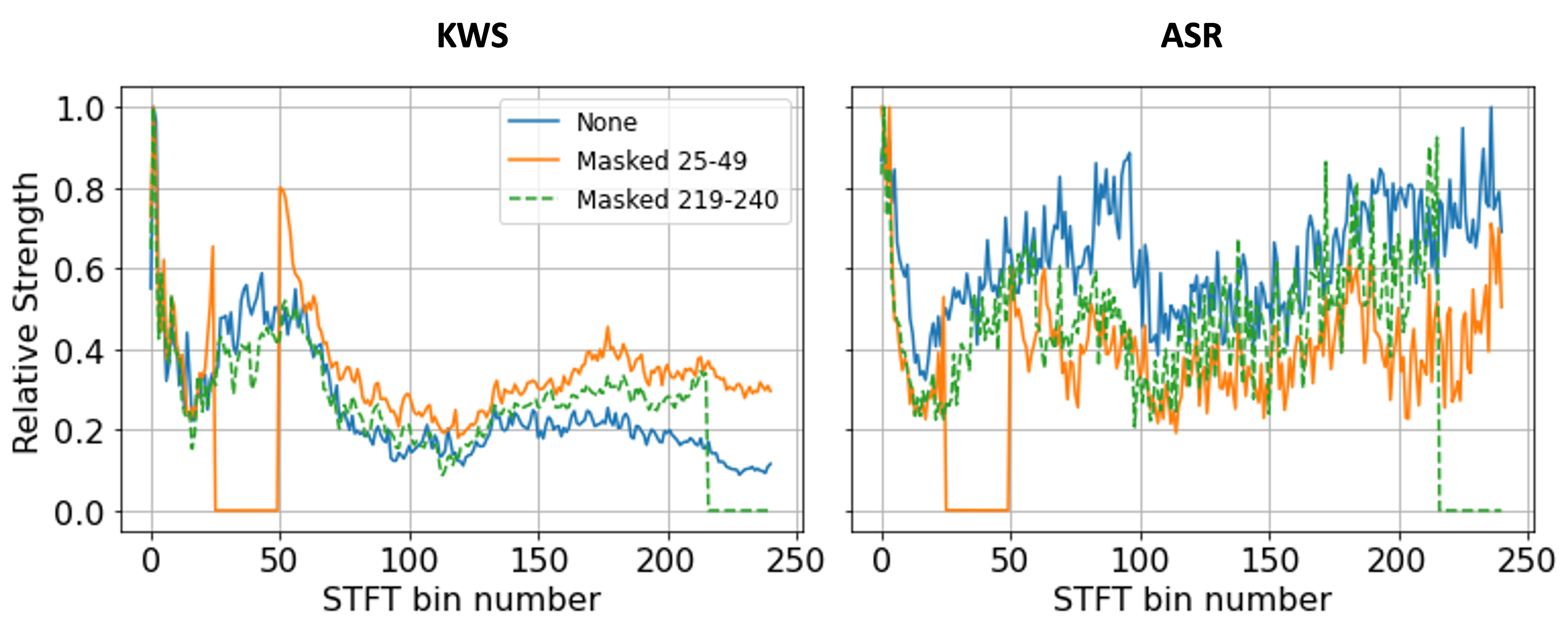}
  \caption{Cumulative strengths when different STFT bins are masked.}
  \label{fig:mask}
\end{figure}

\subsection{Trained Mel}
\label{sec:Mel_filters}

To understand what spectral features has been learned, we visualize the changes in $g_\text{Mel}$ after training on the speech commands dataset in Figure~\ref{fig:visualization}(a)-(b). The following discussion is based on Simple as the classifier with trainable $g_\text{Mel}$ and fixed $g_\text{STFT}$.

Before training, the Mel bins become wider and weaker as the bin number goes up, and they attend to the STFT bins in a log relationship.
After training, each Mel bin focuses less on the STFT bins that they originally attended to. Instead, they attend to more STFT bins than before.
Figure~\ref{fig:visualization}(c) shows the cumulative importance of different STFT bins before and after training. This is calculated by summing all bases in $g_\text{Mel}$ and then normalizing by the largest value. Before training, the $g_\text{Mel}$ extract information mostly from STFT bins 0-30, and the importance gradually decays for the higher STFT bins. After training, the $g_\text{Mel}$ still pay great attention to the first few STFT bins. But STFT bins 10-25 are not as important as before.
Instead, STFT bin numbers above 125 are more important than before. To prove that this finding is not due to the initial states of the $g_\text{Mel}$ leading, we randomly initialized the basis functions, and trained these randomly initialized basis functions end-to-end together with the model. The result is shown as the green lines of Figure~\ref{fig:visualization}(c), which converges to the same pattern as the one initialized as triangular $g_\text{Mel}$ (orange line). We observe a very similar pattern after training on the TIMIT dataset, but $g_\text{Mel}$ tend to pay attention to different frequency bins.

In Table~\ref{tab:mask}, we apply masks to the STFT output to study the effects on the prediction accuracy. Figure~\ref{fig:mask} shows the cumulative STFT bin importance that $g_\text{Mel}$ pays attention to. When STFT bins are masked by 0, the $g_\text{Mel}$ ignores those bins. In the case of KWS, the original accuracy is 42.8\% without applying any mask to the STFT output. When masking bin numbers 25-49, the accuracy increases to 45.6\%. Masking more frequency bins (bin numbers 25-74) makes the accuracy worse than before, which indicates that bin numbers 50-74 contain important information for this task. When bin numbers 216-240 are masked, the model reaches its best accuracy (49.45\%). A similar pattern can be observed for the ASR task.

\subsection{Trained STFT}
Similar to Section~\ref{sec:Mel_filters}, we found that trained $g_\text{STFT}$ also pay more attention to lower frequencies than to higher frequencies.

Figure~\ref{fig:stft_visualization}(a) shows the time-domain STFT filter corresponding to STFT bin number 25 (833 Hz). The original imgainary part of STFT kernel (blue line) is a pure tone containing only one frequency component. The discrete Fourier transform (DFT) in Figure~\ref{fig:stft_visualization}(b) shows that this pure sine wave has only one frequency component (833 Hz). After training, we can see that the amplitude of the sine filter is lower than the original one. While the trained filter maintains the same sine wave shape as its backbone, it superimposes higher frequencies components which can be seen from the DFT analysis in Figure~\ref{fig:stft_visualization}(b). Rather than the original peaks, smaller peaks emerge around it.

Figure~\ref{fig:stft_visualization}(c) shows the the cumulative importance of different frequencies. Similar to $g_\text{Mel}$, all of the frequency components from each STFT filter are summed together and then normalized by the maximum value such that the cumulative importance is in the range of $[0,1]$. The overall trend of the cumulative importance is very similar to Figure~\ref{fig:visualization}(c), where frequencies between 1000Hz and 2000Hz are more important than frequencies above 7000Hz. We observe a very similar pattern for ASR, and are therefore not repeating the same discussion here due to page limits.

\subsection{Corrupted Spectrograms}
Another interesting property of trainable $g_\text{Mel}$ is that it excels when some frequency bins on the STFT output are missing. Here, we keep STFT fixed and train the $g_\text{Mel}$. Table~\ref{tab:mask} shows the model performance when we mask out different STFT frequency bins with the value 0. When STFT bins 216-240 are masked, a simple model with trainable $g_\text{Mel}$ is able to achieve an accuracy of 49.5\% for KWS, which is much higher than 42.8\% (the result when all STFT frequency bins are available). However, the same mask applied when $g_\text{Mel}$ is in a non-trainable state, causes the KWS accuracy to deteriorate to only 18.3\%. A similar pattern can be observed for the ASR task. It is notable that trainable Mel together with missing STFT bins achieves a better result than when all STFT bins are intact.

In Figure~\ref{fig:mask}, we try to understand what is happening when some STFT bins are masked out. When bins 25-49 are missing, the trained $g_\text{Mel}$ pays more attention to bin 24 and 50 and ignore bins 25-49. When bins 216-240 are masked, the trained Mel bases pay attention to bins 200-215 more than before. We believe that the trainable $g_\text{Mel}$ is able to compensate for the loss of information by paying more attention to the remaining available bins. When more bins are masked, for example bins 25-74, the model performance starts to decay. A very similar pattern can be observed for the ASR task.

Based on the experimental results, we believe that trainable basis functions are useful when one of the following two conditions are fulfilled: 1) Model with a limited discriminative power. 2) Input with some missing frequency bins.

\section{Conclusions}
In this paper, we have shown that trainable STFT or Mel basis functions are helpful when the model capability is limited and some frequency bins are missing from the spectrograms. When these conditions are met, the accuracy for KWS can be improved by up to 14.2 ppt., and the PER can be improved by 9.5 ppt. As the model architecture becomes more mature, trainable basis functions become less effective. 
They is still useful under a certain circumstances, and can potentially help us better understand the model's behaviour. The source code of our models is available online\footnote{https://github.com/heungky/trainable-STFT-Mel}.

\begin{table}[]
\caption{Model performance when some frequency bins on STFT are masked by 0. $g_\text{Mel}$ is either trainable or non-trainable but $g_\text{STFT}$ is non-trainable in this experiment.}
\label{tab:mask}
\centering
\begin{tabular}{cccc}
\toprule
\begin{tabular}{@{}c@{}}\textbf{Trainable} \\ $g_\text{Mel}$ \end{tabular}   &
\begin{tabular}{@{}c@{}}\textbf{Masked} \\ \textbf{STFT bins} \end{tabular}   &
\begin{tabular}{@{}c@{}}\textbf{KWS} \\ \textbf{Accuracy} \end{tabular}  &
\begin{tabular}{@{}c@{}}\textbf{ASR} \\ \textbf{PER} \end{tabular}     \\\midrule
\multirow{5}{*}{Yes} &25-49   & 45.6\%      & 40.2   \\
&25-74   & 35.6\%      & 47.3    \\
&216-240 & \textbf{49.5\%}      & \textbf{38.6}    \\
&191-240 & 46.0\%      & 38.7    \\
&None    & 42.8\%      & 41.4  \\\midrule
\multirow{2}{*}{No} &216-240 & 18.3\%      & 50.8  \\
&None & 36.2\%      & 51.0  \\
\bottomrule
\end{tabular}
\end{table}

\section{Acknowledgements}
This work is supported by A*STAR SING-2018-02-0204 and MOE Tier 2 grant MOE2018-T2-2-161.

\bibliographystyle{IEEEtran}

\bibliography{mybib}


\end{document}